\documentclass{article} 

\usepackage{epsfig}

\usepackage{cite}  

\def\NF{N_F}

\def\Poles{{\cal P}oles}

\def\Re{\mbox{Re}}

\def\bom#1{{\mbox{\boldmath $#1$}}}

\def\e{\epsilon}

\def\d{\hbox{d}}

\begin{document} 

\unitlength1cm 

\begin{titlepage} 

\vspace*{-1cm} 

\begin{flushright} 
ZU--TH 01/05\\
IPPP/05/01\\
DCPT/05/02\\
hep-ph/0501291\\
January 2005
\end{flushright} 

\vskip 2.5cm

\begin{center} 
{\Large\bf Quark-Gluon Antenna Functions from Neutralino Decay}
\vskip 1.cm 
{\large  A.~Gehrmann--De Ridder}$^{a}$, {\large  T.~Gehrmann}$^{b}$ 
and {\large E.W.N.~Glover}$^{c}$ 
\vskip .7cm 
{\it $^a$ Institute for Theoretical Physics, ETH, CH-8093 Z\"urich,
Switzerland} 
\vskip .4cm 
{\it $^b$ Institut f\"ur Theoretische Physik, Universit\"at Z\"urich,
Winterthurerstrasse 190,\\ CH-8057 Z\"urich, Switzerland} 
\vskip .4cm 
{\it $^c$ Institute for Particle Physics Phenomenology, University of Durham,
South Road,\\ Durham DH1 3LE, England} 
\end{center} 
\vskip 2.6cm 
\begin{abstract} 
The computation of exclusive QCD jet observables at higher orders 
requires a method for the subtraction of infrared singular
configurations arising from multiple radiation of real partons. 
One commonly used method at next-to-leading order (NLO) is based on 
the antenna factorization of colour-ordered matrix elements, and uses 
antenna functions to subtract the real radiation singularities. Up to 
now, NLO antenna functions could be derived in a systematic manner
only for hard
quark-antiquark pairs, while the gluon-gluon and quark-gluon 
antenna functions were constructed from their limiting 
behaviour. In this paper, we show that antenna functions for hard 
quark-gluon  pairs can be systematically 
derived from an effective Lagrangian describing heavy neutralino decay. 
The infrared structure of the colour-ordered neutralino decay matrix 
elements at NLO and NNLO is shown to agree with the structure observed 
for parton radiation off a quark-gluon antenna. 
\end{abstract} 
\vfill 

\end{titlepage} 

\newpage 

\renewcommand{\theequation}{\mbox{\arabic{section}.\arabic{equation}}}

\section{Introduction}
\setcounter{equation}{0}

Experimental measurements of jet production observables are among the 
most sensitive tests of the theory of Quantum Chromodynamics (QCD), 
and yield very accurate determinations of QCD parameters~\cite{dissertori},
especially of the strong coupling constant $\alpha_s$. At present, 
the precision of many of these determinations is limited not by the 
quality of the experimental data, but 
by the error on the theoretical (next-to-leading order, NLO) 
calculations used for the extraction of 
the QCD parameters. To 
improve upon this situation, an extension of the theoretical calculations to 
next-to-next-to-leading order (NNLO) is therefore mandatory.

In the recent past, many ingredients to NNLO calculations of collider
 observables have been derived,
including the universal three-loop QCD splitting 
functions~\cite{mvv} which
governs the evolution of parton distribution functions at NNLO. 
The massless
two-loop $2\to 2$ and $1 \to 3$ matrix elements relevant to 
NNLO jet production 
have been computed~\cite{twol} using several innovative methods~\cite{twolmeth},
and are now available for all processes of phenomenological relevance.
The one-loop corrections to $2\to 3$ and $1\to 4$ matrix elements 
have been known for longer and form part of NLO calculations of 
the respective multi-jet observables~\cite{nlomult,cullen}. 
These NLO matrix elements naturally
contribute to NNLO jet observables of lower multiplicity if one of the 
partons involved becomes soft or collinear~\cite{onelstr}. 
In these cases, the infrared 
singular parts of the matrix elements need to be extracted and integrated 
over the phase space appropriate to the unresolved configuration 
to make the infrared pole structure explicit. Methods for the 
extraction of soft and collinear limits of one-loop matrix elements are  
worked out in detail in the literature~\cite{onel}. As a final ingredient,
contributions from the tree level $2\to 4$ and $1\to 5$ processes also
contribute to ($2\to 2$)- and ($1 \to 3$)-type jet observables at NNLO. These 
contain double real radiation singularities corresponding to two 
partons becoming simultaneously soft and/or 
collinear~\cite{campbell,campbellandother}. 
To determine the
contribution to NNLO jet observables from these configurations, one has to 
find two-parton subtraction terms which coincide with the full matrix element 
and are still sufficiently simple to be integrated analytically in order 
to cancel  their  infrared pole structure with the two-loop virtual and 
the one-loop single-unresolved contributions. Several methods have been 
proposed recently to accomplish this task~\cite{nnlosub}. Up to now, only 
one method has been fully worked through for an observable of physical
interest: using the sector decomposition algorithm~\cite{secdec,ggh} to 
analytically decompose both phase space and loop integrals into their 
Laurent expansion in dimensional regularization, and subsequent numerical 
computation of the coefficients of this expansion, results were 
obtained for $e^+e^- \to 2j$~\cite{babis2j} 
and $pp \to H+X$~\cite{babishiggs} at NNLO. In contrast to all other
approaches, in the  sector decomposition method one does not have to 
integrate the subtraction term analytically. 

In~\cite{our2j}, we described the construction of NNLO subtraction terms 
for $e^+e^- \to 2j$ based on full four-parton tree-level 
and three-parton one-loop matrix elements, which can be integrated
analytically over the appropriate phase spaces~\cite{ggh}. 
Subtraction terms derived from full matrix elements can be viewed as 
antenna functions, encapsulating all singular limits due to
unresolved  
partonic emission between two colour-connected hard
partons~\cite{cullen,ant}. 
In particular, process-independent 
antenna functions describing 
arbitrary QCD multiparticle processes can be directly related to three-parton 
matrix elements at NLO (one unresolved parton radiating between two 
colour-connected hard partons)
and four-parton matrix elements at NNLO (two unresolved 
partons radiating between  two 
colour-connected hard partons).

Up to now, antenna subtraction terms (at NLO) were obtained 
by construction (i.e.\ by inspecting all limits they had to contain), 
in part from the full matrix elements, and in part by using supersymmetric
(SUSY) relations between matrix elements containing  fermions and 
bosons~\cite{cullen}. A systematic procedure to derive antenna functions 
at NLO and beyond is not available up to now: this paper aims to contribute to 
such a formalism by showing that quark-gluon antenna functions can be 
derived systematically from physical matrix elements obtained from
an effective Lagrangian. 

The NNLO subtraction terms
 derived from four-parton matrix elements with a hard quark-antiquark
pair 
in~\cite{our2j} were used subsequently~\cite{our3j} 
to compute the $\alpha_s^3 C_F^3$--correction to $e^+e^- \to 3j$ 
at NNLO. To extend this calculation to the remaining colour factors,
further subtraction terms must be derived. In particular, 
the subtraction terms of~\cite{our2j} are sufficient only for  
processes where unresolved partons are radiated from hard
quark-antiquark pairs:
they form the quark-antiquark antenna functions at NNLO. 
Besides quark-antiquark antennae,  $e^+e^- \to 3j$ also contains 
radiation from hard quark-gluon pairs, the quark-gluon
antenna function. In the spirit of~\cite{our2j}, it should be 
possible to extract these antenna functions from the matrix elements 
appearing in the NLO and NNLO corrections
to a physical one-particle decay process 
yielding a quark-gluon final state at leading order. It is the
purpose of this letter to show that such a process can be described 
by an appropriate colour ordering of the decay of a 
massive neutralino into a massless gluino and a gluon, and to derive the 
resulting quark-gluon antenna subtraction terms at NLO and NNLO.

\section{Effective Lagrangian and Feynman rules}
\setcounter{equation}{0}

To obtain the correct quantum numbers for a quark-gluon antenna function,
one has to consider the decay of an off-shell
spin-$1/2$ particle into an on-shell spin-$1/2$ particle (massless quark)
and an on-shell spin-$1$ particle (gluon). Since the final state 
quark is in the triplet representation of $SU(3)$, while 
the gluon is in the octet representation, this implies that the 
initial state spin-$1/2$ off-shell particle  should also be in the 
triplet representation. $SU(3)$ gauge invariance does however forbid 
external off-shell states. 

In the colour-ordered formulation of QCD tree-level 
amplitudes~\cite{colord,ddm}, 
one decouples the colour quantum numbers of the partons from their 
Lorentz and Dirac structure. Using this formulation, one can in particular 
represent a parton in the adjoint representation as superposition of 
two partons (with identical momenta) in the fundamental representation.
It is thus possible to construct the colour ordered quark-gluon antenna 
functions from the $SU(3)$ gauge-invariant decay of an off-shell spin-$1/2$
{\it singlet} state (neutralino)
into a spin-$1/2$ octet state (massless gluino) and 
a spin-$1$ octet state (gluon), as we shall show below.

This decay process occurs in the minimal supersymmetric standard model 
(MSSM,\cite{mssm}), where it is mediated through a loop involving
supersymmetric particles. For the purpose of this study, it is sufficient 
to describe this process through an effective Lagrangian, whose parameters 
are obtained by integrating out the virtual particles in the loop. 
In the context of the electroweak sector of the MSSM, this effective 
Lagrangian was first derived by Haber and Wyler~\cite{hw}, to describe 
heavy neutralino decay into a light neutralino and a photon. Its
generalization to neutralino decay into gluino and gluon is 
straightforward:
\begin{equation}
{\cal L}_{{\rm int}} = i \eta \overline{\psi}^a_{\tilde{g}} \sigma^{\mu\nu} 
\psi_{\tilde{\chi}} F_{\mu\nu}^a  + ({\rm h.c.})\; ,
\label{eq:hw}
\end{equation}
which couples a gluino ($\overline{\psi}^a_{\tilde{g}}$) and a 
neutralino ($\psi_{\tilde{\chi}}$)  to the QCD field strength tensor 
$F_{\mu\nu}^a$. The coupling $\eta$ has inverse mass dimension, and 
the commutator of the $\gamma$-matrices is
$$\sigma^{\mu\nu} = \frac{i}{2} \left[\gamma^\mu,\gamma^\nu \right]\;.$$
It should be noted that this process was discussed previously in the 
literature in~\cite{tata}, where however no effective
Lagrangian was stated. 

The Feynman rules following from the  
Lagrangian (\ref{eq:hw}) are
\begin{eqnarray}
\parbox{6cm}{\epsfig{file=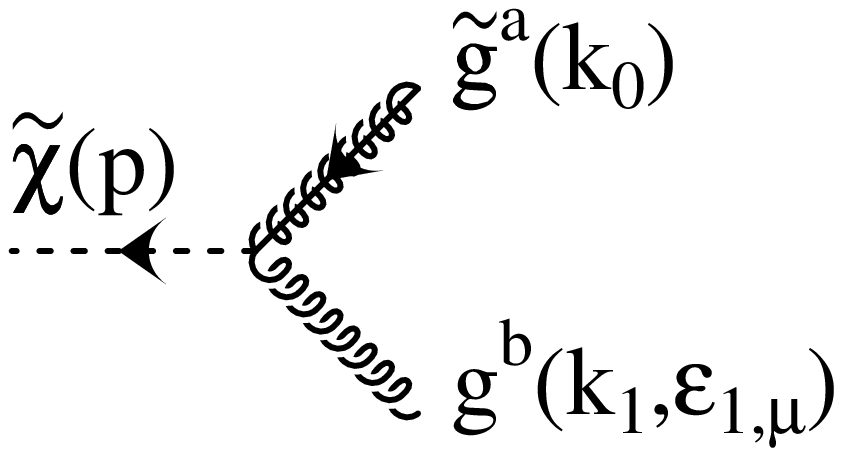,width=5.8cm}} &=& - i \eta 
\delta^{ab} \sigma_{\mu\nu}
k_{1}^{\nu}\;, \nonumber \\[6mm]
\parbox{6cm}{\epsfig{file=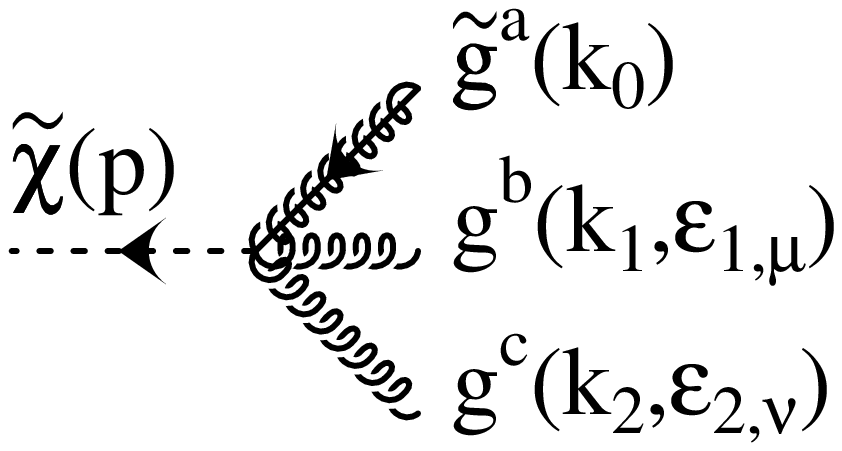,width=5.8cm}} &=& - g_s \eta f^{abc} 
\sigma_{\mu\nu}\;,
\end{eqnarray}
where the arrow indicates the direction of fermion flow. It 
should be noted that Majorana 
particles also have a fermion flow direction, which does however not 
coincide with the fermion number flow. The
momenta are always incoming.

Besides these 
Feynman rules and the standard QCD Feynman rules, one needs 
moreover the Feynman rule for the gluon-gluino-gluino coupling~\cite{rosiek}
\begin{equation}
\parbox{6cm}{\epsfig{file=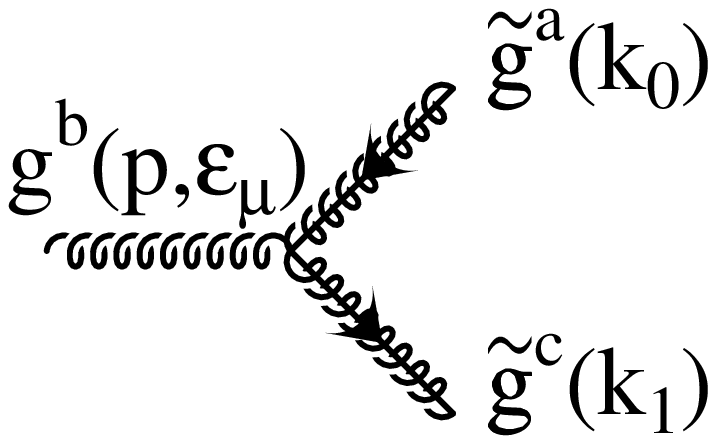,width=5.8cm}} = -  g_s f^{abc} \gamma^\mu\;.
\end{equation}

The effective coupling $\eta$ can be computed in the MSSM, it was discussed 
in~\cite{hw,tata}. In the present context, its value is irrelevant, but we 
do have to take into account that $\eta$ is renormalized at one loop. Its 
QCD renormalization constant reads
\begin{equation}
Z_\eta = 1 - \frac{\alpha_s}{2\pi} \, \frac{1}{\e} \, \left(
\frac{\beta_0}{2} + \frac{3N}{4} \right) + {\cal O} (\alpha_s^2)\;,
\label{eq:Zeta}
\end{equation}
with
\begin{equation}
\beta_0 = \frac{11 N - 2 N_F}{6}\;.
\label{eq:qcdbeta}
\end{equation}
In SUSY QCD, both $Z_{\eta}$ and $\beta_0$ are modified. Throughout this paper,
we systematically ignore the SUSY QCD corrections and restrict ourselves to 
the subclass of contributions which preserve the QCD renormalizations
(\ref{eq:Zeta}) and (\ref{eq:qcdbeta}). 

\section{Colour-ordered amplitudes in neutralino decay}
\setcounter{equation}{0}

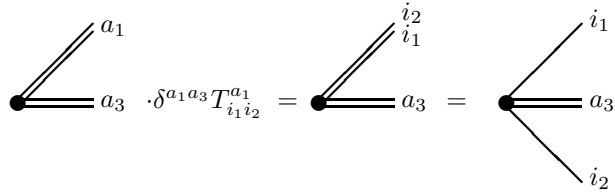
\begin{figure}[t]
\begin{center}
\parbox{8cm}{\begin{picture}(8,3)
\thicklines
\put(0.0,1.5){\line(1,1){1.0}}
\put(0.0,1.5){\line(1,0){1.0}}
\put(0.0,1.4){\line(1,1){1.0}}
\put(0.0,1.4){\line(1,0){1.0}}
\put(0.0,1.45){\circle*{0.2}}
\put(1.1,1.45){\makebox(0,0)[l]{$a_3$}}
\put(1.1,2.45){\makebox(0,0)[l]{$a_1$}}
\put(1.7,1.45){\makebox(0,0)[l]{$\cdot \delta^{a_1a_3}T^{a_1}_{i_1i_2}\; =$}}
\put(4.0,1.5){\line(1,1){1.0}}
\put(4.0,1.5){\line(1,0){1.0}}
\put(4.0,1.4){\line(1,1){1.0}}
\put(4.0,1.4){\line(1,0){1.0}}
\put(4.0,1.45){\circle*{0.2}}
\put(5.1,1.45){\makebox(0,0)[l]{$a_3$}}
\put(5.1,2.35){\makebox(0,0)[l]{$i_1$}}
\put(5.1,2.65){\makebox(0,0)[l]{$i_2$}}
\put(5.7,1.45){\makebox(0,0)[l]{$=$}}
\put(6.5,1.5){\line(1,1){1.0}}
\put(6.5,1.5){\line(1,0){1.0}}
\put(6.5,1.4){\line(1,-1){1.0}}
\put(6.5,1.4){\line(1,0){1.0}}
\put(6.5,1.45){\circle*{0.2}}
\put(7.6,1.45){\makebox(0,0)[l]{$a_3$}}
\put(7.6,0.45){\makebox(0,0)[l]{$i_2$}}
\put(7.6,2.55){\makebox(0,0)[l]{$i_1$}}
\end{picture}}
\end{center}
\caption{Colour flow contained in tree level 
decay $\tilde \chi \to \tilde g g$. Double (single) lines denote adjoint
(fundamental) colour indices.}
\label{fig:ant0}
\end{figure}

The basic process for the decay of a neutralino into 
a gluino plus partons is 
$\tilde{\chi}(q) \to \tilde g(p_1) g(p_3)$. Its amplitude reads
\begin{displaymath}
i \eta \delta^{a_1 a_3} M_{\tilde g g}^0 (p_1,p_3) \;.
\end{displaymath}
To display the colour-ordered structure of this amplitude, and to 
illustrate the relation to quark-gluon amplitudes, we multiply it with 
$\sqrt{2} T^{a_1}_{i_1 i_2}$~\cite{ddm}. 
In the squared amplitude, this factor
corresponds to inserting unity since
\begin{displaymath}
\sqrt{2} T^{a_1}_{i_1 i_2} \sqrt{2} T^{a_1'}_{i_2 i_1} = \delta^{a_1a_1'}\;.
\end{displaymath}
The resulting amplitude is then 
\begin{equation}
{\cal M}^0_{\tilde g_1 g_3} 
= i \eta\, \sqrt{2}\,
T_{i_1i_2}^{a_3} M_{\tilde g g}^0 (p_1,p_3) \;.
\label{eq:D20}
\end{equation}
From this structure, it can be seen that the amplitude contains two 
colour connected (hard) partons and therefore two antennae:
\begin{enumerate}
\item A quark-gluon antenna, with quark momentum $p_1$, quark 
colour index $i_1$, gluon momentum $p_3$ and gluon colour index 
$a_3$. 
\item An antiquark-gluon antenna, with antiquark momentum $p_1$, 
antiquark colour index $i_2$, gluon momentum $p_3$ and gluon colour 
index $a_3$.
\end{enumerate}
This derivation is displayed pictorially in Figure~\ref{fig:ant0}. It becomes 
evident that the Majorana nature of the gluino allows it to represent 
both a quark and an anti-quark. 

The squared matrix element is
\begin{equation}
{\cal T}^0_{\tilde g g}(q^2) \equiv |{\cal M}^0_{\tilde g_1 g_3}|^2 
= \eta^2 \left(N^2-1\right) 
|M_{\tilde g g}^0 (p_1,p_3)|^2  = 
4 \left(N^2-1\right)\, \eta^2(1-\e)(q^2)^2 \; .
\end{equation}
${\cal T}^0_{\tilde g g}(q^2) $  serves as normalization for 
antenna functions obtained from higher order 
corrections to this matrix element.

To demonstrate the cancellation of infrared divergences at NLO, we
compute the renormalized one-loop QCD correction to the 
$\tilde{\chi}(q) \to \tilde g(p_1) g(p_3)$ decay,
\begin{eqnarray}
{\cal T}^1_{\tilde g g}(q^2) &\equiv& 
2\mbox{Re}|{\cal M}^0_{\tilde g_1 g_3}{\cal M}^{1,*}_{\tilde g_1 g_3}
|\nonumber \\
&=&\left(\frac{\alpha_s}{2\pi}\right)
2 (q^2)^{-\e}\, 
{\cal T}^0_{\tilde g g}(q^2) 
\Bigg\{ N \Bigg[ -\frac{1}{\e^2} - \frac{5}{3\e}  + 
\frac{7\pi^2}{12} 
+ \left( -1 + \frac{7}{3}\zeta_3 \right)
\e 
+ \left( - 3
- \frac{73\pi^4}{1440} \right) \e^2  \Bigg] \nonumber \\
&& \hspace{3.5cm}+ 
\frac{N_F}{6\e} + {\cal O}(\e^3)\Bigg\} \;.
\end{eqnarray}
The infrared poles of this one-loop correction can be expressed in terms 
of the infrared singularity operator\cite{catani}
\begin{equation}
\bom{I}_{qg}^{(1)}(\epsilon,q^2)
= - \frac{e^{\epsilon\gamma}}{2\Gamma(1-\epsilon)} \left[
N\,
\left( \frac{1}{\e^2}+\frac{3}{4\epsilon}+\frac{\beta_0}{2N\e}\right)
\,\left(-q^2\right)^{-\e}\,\right]\;  
\label{eq:I1}
\end{equation}
as
\begin{equation}
\Poles\left( {\cal T}^1_{\tilde g g}(q^2)  \right) = 
\left(\frac{\alpha_s}{2\pi}\right)\;
4 \Re \bom{I}_{qg}^{(1)}(\epsilon,q^2)\;{\cal T}^0_{\tilde g g}(q^2)\;.
\label{eq:poles1l}
\end{equation}
This expression has to be compared to the 
$2 \Re \bom{I}_{q\bar q}^{(1)}(\epsilon,q^2)$, which is obtained in the 
decay of a virtual photon into a quark-antiquark pair 
$\gamma^* \to q \bar q$ at one loop~\cite{our2j}. The factor $4$ in 
(\ref{eq:poles1l}) appears since the leading order process 
$\tilde{\chi} \to \tilde g g$ contains two distinct 
quark-gluon antennae, 
in contrast to the single quark-antiquark antenna contained in 
$\gamma^* \to q \bar q$.

\section{NLO antenna functions}
\setcounter{equation}{0}
Two different emissions off a quark-gluon pair appear at NLO:
either the emission of an additional gluon or the splitting of the 
gluon into a quark-antiquark pair. In the context of the 
neutralino decay, these correspond to the tree level 
processes $\tilde{\chi} \to \tilde{g} gg$ and $\tilde{\chi}\to \tilde{g} q 
\bar q$.

The tree level amplitude for $\tilde{\chi}(q) \to \tilde{g}(p_1) 
g (p_3) g(p_4)$
contains only a single colour structure, $f^{a_1a_3a_4}$. In order to
relate this colour structure to the colour-ordered 
quark-gluon antennae,
 we multiply with $\sqrt{2}T^{a_1}_{i_1i_2}$:
\begin{equation}
{\cal M}^0_{\tilde g_1 g_3 g_4} 
= i \eta g (-i\sqrt{2}) \left[\left(T^{a_3}T^{a_4}\right)_{i_1i_2}
- \left(T^{a_4}T^{a_3}\right)_{i_1i_2} \right]
 M_{\tilde g g g}^0 (p_1,p_3,p_4) \;,
\end{equation}
showing that the two colour-ordered amplitudes in this matrix 
element (corresponding to the two different orderings of the gluons 
along the quark-antiquark $i_1i_2$ colour line)
are equivalent to each other up to an overall sign
because of the identical momenta of quark and antiquark. Squaring the
matrix element and dividing by a symmetry factor to account for 
identical gluons in the final state yields
\begin{equation}
\frac{1}{2}\,|{\cal M}^0_{\tilde g_1 g_3 g_4}|^2 = \eta^2 g^2 \,
\left(N^2-1\right) N 
\, \frac{1}{2} |M_{\tilde g g g}^0 (p_1,p_3,p_4)|^2 \;,
\end{equation}
with
\begin{eqnarray}
\frac{1}{2}\,|M_{\tilde g g g}^0 (p_1,p_3,p_4)|^2  &=& 
4\, (1-\e)\, \Bigg(
\frac{2 s_{134}^2 s_{14}}{ s_{13}  s_{34}}
+\frac{2 s_{134}^2 s_{13}}{ s_{14}  s_{34}}
+ \frac{(1-\e)\,s_{134} s_{34}}{s_{13}} 
+ \frac{(1-\e)\,s_{134} s_{34}}{s_{14}}\nonumber \\&&
+ \frac{2s_{13}s_{14}}{ s_{34}} + 6  s_{134} + (1-\e) 
\left(s_{13}+s_{14} \right)
\Bigg) - 8 s_{134} \; .
\end{eqnarray}

The behaviour of this matrix element in the kinematical limits where one 
parton becomes unresolved is as follows:
\begin{enumerate}
\item Collinear limits:
\begin{eqnarray}
\frac{1}{2}\,|{\cal M}^0_{\tilde g_1 g_3 g_4}|^2 
&\stackrel{\tilde g_1 \parallel
g_3}{\longrightarrow}& \left({4\pi\alpha_s}\right)\;
{\cal T}^0_{\tilde g g}(s_{134}) \frac{1}{s_{13}}\,
N \,  P_{q\to qg}(z)\;, \nonumber \\
\frac{1}{2}\,|{\cal M}^0_{\tilde g_1 g_3 g_4}|^2
 &\stackrel{\tilde g_1 \parallel
g_4}{\longrightarrow}& \left({4\pi\alpha_s}\right)\;
{\cal T}^0_{\tilde g g}(s_{134}) \frac{1}{s_{14}}\,
N \,  P_{q\to qg}(z)\;, \nonumber \\
\frac{1}{2}\,|{\cal M}^0_{\tilde g_1 g_3 g_4}|^2
 &\stackrel{g_3 \parallel
g_4}{\longrightarrow}& \left({4\pi\alpha_s}\right)\;
{\cal T}^0_{\tilde g g}(s_{134}) \frac{1}{s_{34}}\,
N \,  P_{g\to gg}(z) \;,
\end{eqnarray}
with $z$ being the momentum fraction of one of the collinear partons and
the splitting functions 
\begin{displaymath}
P_{q\to qg}(z) = \frac{1+z^2}{1-z} - \e (1-z)\, ,
 \qquad P_{g\to gg}(z) = 2\left[\frac{z}{1-z}+ \frac{1-z}{z}+z(1-z)\right]\;.
\end{displaymath}
\item Soft limits:
\begin{eqnarray}
\frac{1}{2}\,|{\cal M}^0_{\tilde g_1 g_3 g_4}|^2 &\stackrel{g_3 \to
  0}{\longrightarrow}& \left({4\pi\alpha_s}\right)\;
{\cal T}^0_{\tilde g g}(s_{134})\,  N\, 
\frac{2s_{14}}{s_{13}s_{34}}\; , \nonumber \\
\frac{1}{2}\,|{\cal M}^0_{\tilde g_1 g_3 g_4}|^2 &\stackrel{g_4 \to 
0}{\longrightarrow}& \left({4\pi\alpha_s}\right)\;
{\cal T}^0_{\tilde g g}(s_{134})\,  N \,
\frac{2s_{13}}{s_{14}s_{34}}\;. 
\end{eqnarray}
\end{enumerate}
Comparing these limits to the limits of colour-ordered QCD matrix elements, 
one observes that the collinear $q\to qg$ limit contains a colour factor
$N/2$ in QCD, while the collinear $\tilde g \to \tilde g g$ limit 
derived here contains a colour factor $N$. This is precisely what was expected 
from the discussion in the previous section, since the neutralino 
decay matrix element considered here contains both a quark-gluon and 
an antiquark-gluon antenna. On the other hand, the collinear $g\to gg$ 
limit appears here with the same colour factor as in colour ordered 
QCD matrix elements, indicating that the   collinear $g\to gg$ limit is 
to be split between both antenna functions, as 
discussed in~\cite{cullen,ant}. Finally, the matrix element derived here 
contains two soft limits with the soft eikonal factors as expected in QCD, 
again reflecting the presence of two antennae.

Integration over the dipole phase space~\cite{ggh} yields 
\begin{eqnarray}
{\cal T}^1_{\tilde g gg}(q^2) 
&\equiv&
\int \d \Phi_{D,\tilde g gg}\;\frac{1}{2}|{\cal M}^0_{\tilde g_1 g_3 g_4}|^2
\nonumber \\
&=&  \left(\frac{\alpha_s}{2\pi}\right)\; N \;{\cal T}^0_{\tilde g g}(q^2)
\left(q^2 \right)^{-\e} 
  \Bigg[
\frac{2}{\e^2} + \frac{10}{3\e} + \frac{34}{3} - 
\frac{7\pi^2}{6} 
+ \left( \frac{209}{6} - \frac{35\pi^2}{18} - \frac{50}{3}\zeta_3 \right)
\e \nonumber \\
&& \hspace{2cm}
+ \left( \frac{421}{4}
          - \frac{119\pi^2}{18}
          - \frac{250}{9}\zeta_3
          - \frac{71\pi^4}{720} \right) \e^2  + {\cal O}(\e^3) \Bigg] \; .
\end{eqnarray}

The tree level amplitude for $\tilde{\chi}(q) \to \tilde{g}(p_1)
q(p_3) \bar q(p_4)$
contains only a single colour structure $T^{a_1}_{i_3 i_4}$, which is 
again contracted with $\sqrt{2} T^{a_1}_{i_1i_2}$ 
\begin{equation}
{\cal M}^0_{\tilde{g}_1 q_3 \bar q_4} 
= i \eta g \frac{1}{\sqrt{2}} \, \left( 
\delta_{i_1i_4} \delta_{i_3i_2} - \frac{1}{N}
\delta_{i_1i_2} \delta_{i_3i_4} 
\right) M_{\tilde{g} q \bar q}^0 (p_1,p_3,p_4)
\;,
\end{equation}
yielding 
\begin{equation}
|{\cal M}^0_{\tilde{g}_1 q_3 \bar q_4}|^2 = \eta^2 
g^2 \frac{N^2-1}{2} 
|M_{\tilde{g} q \bar q}^0 (p_1,p_3,p_4)|^2 \;,
\end{equation}
with 
\begin{equation}
|M_{\tilde{g} q \bar q}^0 (p_1,p_3,p_4)|^2 = 
4 \left( 1-\e \right) \left(2 \frac{\left(s_{13}+s_{14}\right)^2}{s_{34}}
+ 2 \left(s_{13}+s_{14}\right)\, \right) - 16 \frac{s_{13}s_{14}}{s_{34}}\;.
\end{equation}
The only singular configuration contained in this matrix element is 
the collinear quark-antiquark limit, which is as follows:
\begin{equation}
|{\cal M}^0_{\tilde g_1 q_3 \bar q_4}|^2 \stackrel{q_3 \parallel
\bar q_4}{\longrightarrow} \left({4\pi\alpha_s}\right)
\;
{\cal T}^0_{\tilde g g}(s_{134}) \frac{1}{s_{34}}\,
  P_{g\to q\bar q}(z) \;,
\end{equation}
with the collinear splitting function
\begin{displaymath}
P_{g\to q\bar q}(z) = 1 - \frac{2z(1-z)}{1-\e}\;.
\end{displaymath}

Integration over the dipole phase space~\cite{ggh} and summing 
over final state quark flavours yields 
\begin{eqnarray}
{\cal T}^1_{\tilde g q \bar q} (q^2) &\equiv&
\int \d \Phi_{D,\tilde g q \bar q}\; \sum_q\, 
|{\cal M}^0_{\tilde{g}_1 q_3 \bar q_4}|^2 
 \nonumber \\
&=&  \left(\frac{\alpha_s}{2\pi}\right)\;
 N_F\, {\cal T}^0_{\tilde g g} (q^2) \left(q^2 \right)^{-\e} 
  \Bigg[
- \frac{1}{3\e} -1   
+ \left( -3 + \frac{7\pi^2}{36} \right)
\e + \left( -9 
          + \frac{7\pi^2}{12}
          + \frac{25}{9}\zeta_3
\right) \e^2  + {\cal O}(\e^3) \Bigg] \; .\nonumber \\
\end{eqnarray}

Summing over both three parton final states, we find
\begin{equation}
\Poles\left( {\cal T}^1_{\tilde g g g}(q^2) + 
{\cal T}^1_{\tilde g q \bar q}(q^2)  \right) =
-\left(\frac{\alpha_s}{2\pi}\right)\;4 \Re \bom{I}_{qg}^{(1)}(\epsilon,q^2)\;{\cal T}^0_{\tilde g g}(q^2)\;,
\label{eq:polestree}
\end{equation}
such that the NLO corrected neutralino decay rate into gluino plus partons 
is finite:
\begin{equation}
\Poles\left( {\cal T}^1_{\tilde g g }(q^2) \right) + 
\Poles\left( {\cal T}^1_{\tilde g g g}(q^2) + 
{\cal T}^1_{\tilde g q \bar q}(q^2)  \right) = 0\;.
\end{equation}
It has to be emphasised in this context that we considered here only the QCD 
corrections to the neutralino decay, not the SUSY QCD corrections. 
At NLO, inclusion of SUSY QCD
corrections would both modify the renormalization (\ref{eq:Zeta})
of the effective coupling $\eta$, and include a real radiation contribution 
from the three gluino final state (which has however the same singularity 
structure as the $\tilde g q \bar q$ final state). The omission of these 
corrections is deliberate, since we want to derive the QCD
quark-gluon antenna functions. In the following section, we demonstrate
that the NNLO infrared singularity structure of QCD quark-gluon antenna 
functions is 
reproduced correctly by the  QCD corrections to neutralino decay into 
gluino plus partons.

\section{Structure of NNLO antenna functions}
\setcounter{equation}{0}

In the NNLO calculation of jet observables, two different types of antenna 
functions are required: (a) the one-loop correction to the three-parton antenna 
functions which appeared at NLO in tree-level form, and (b) 
the tree-level 
four-parton antenna functions. In this section, we present all neutralino 
decay matrix elements needed for the derivation of these antenna functions, 
and demonstrate that these matrix elements contain the same  infrared 
singularities as processes involving final state emission off a quark-gluon 
antenna. 

The renormalized 
one-loop corrections to the three-parton antenna functions have the 
same colour structure as their tree level counterparts listed above. 
In their computation, closed gluino loops are omitted, since these 
form part of the SUSY QCD corrections. Consequently, renormalization of the 
coupling constant is done using the QCD $\beta$-function.
To expose the infrared structure of the resulting one-loop matrix elements, 
they are 
integrated over the corresponding dipole phase space~\cite{ggh}, 
yielding
\begin{eqnarray}
{\cal T}^2_{\tilde g gg}(q^2) 
&\equiv&
\int \d \Phi_{D,\tilde g gg}\;\frac{1}{2}
2 \mbox{Re}\left({\cal M}^0_{\tilde g_1 g_3 g_4}
{\cal M}^{1,*}_{\tilde g_1 g_3 g_4}\right)
\nonumber \\
&=& \left(\frac{\alpha_s}{2\pi}\right)^2\; {\cal T}^0_{\tilde g g}(q^2)
\left(q^2 \right)^{-2\e} 
  \Bigg[ N^2 \Bigg( 
-\frac{9}{2\e^4}  - \frac{56}{3\e^3} + \frac{1}{\e^2} 
\left(-\frac{1835}{36} + \frac{71\pi^2}{12} \right) \nonumber \\
&&+ \frac{1}{\e} \left(
 - \frac{20977}{108}
          + \frac{209\pi^2}{12}
          + 72\zeta_3 \right) 
+ \left(  - \frac{19499}{27}
          + \frac{4195\pi^2}{72}
          + \frac{695}{3}\zeta_3
          - \frac{995\pi^4}{720} 
\right)\Bigg) \nonumber \\
&& + N N_F \Bigg( \frac{4}{3\e^3} + \frac{20}{9\e^2} 
+ \frac{1}{\e} \left(\frac{275}{36}
          - \frac{7\pi^2}{9}
\right) + \left(\frac{287}{12}
          - \frac{35\pi^2}{27}
          - \frac{100}{9}\zeta_3
\right)
\Bigg)
+ {\cal O}(\e) \Bigg] \; , \\
{\cal T}^2_{\tilde g q\bar q}(q^2) 
&\equiv&
\int \d \Phi_{D,\tilde g q\bar q}\;
2 \mbox{Re}\left({\cal M}^0_{\tilde g_1 q_3 \bar q_4}
{\cal M}^{1,*}_{\tilde g_1 q_3 \bar q_4}\right)
\nonumber \\
&=&  \left(\frac{\alpha_s}{2\pi}\right)^2\;{\cal T}^0_{\tilde g g}(q^2)
\left(q^2 \right)^{-2\e} 
  \Bigg[ N N_F \Bigg( 
\frac{2}{3\e^3}  + \frac{67}{18\e^2} + \frac{1}{\e} 
\left(\frac{326}{27} - \frac{8\pi^2}{9} \right)
+  \left(
  \frac{9215}{216}
          - \frac{275\pi^2}{72}
          - \frac{94}{9}\zeta_3 \right) 
\Bigg) \nonumber \\
&& + \frac{N_F}{N} \Bigg( -\frac{1}{6\e^3} - \frac{35}{36\e^2} 
+ \frac{1}{\e} \left(-\frac{509}{108}
          + \frac{\pi^2}{4}
\right) + \left(-\frac{1670}{81}
          + \frac{35\pi^2}{24}
          + \frac{31}{9}\zeta_3
\right)
\Bigg)\nonumber \\
&& + N_F^2 \Bigg( -\frac{1}{9\e^2} + \left(\frac{91}{81}
          - \frac{\pi^2}{27}
\right)
\Bigg)
+ {\cal O}(\e) \Bigg] \; .
\end{eqnarray}

Two different four-parton final states appear in the quark-gluon antenna 
functions at NNLO: $qggg$ and $qq'\bar q' g$. The Lorentz and Dirac 
structure of these antenna functions is contained in the neutralino 
decay processes $\tilde \chi \to \tilde g ggg$ and $\tilde \chi \to 
\tilde g q\bar q g$. In contrast to the 
tree level three-parton neutralino decay matrix elements, 
which contained only one non-trivial colour 
ordering each, these four-parton matrix elements both contain several 
colour-orderings.

To expose the colour-ordered subamplitudes contributing to
$\tilde{\chi}(q) \to \tilde{g}(p_1) 
g (p_3) g(p_4) g(p_5)$, we again contract the amplitude with 
$\sqrt{2} T^{a_1}_{i_1i_2}$.
The amplitude can then be expressed as sum over the permutations 
of the gluon colour indices:
\begin{equation}
{\cal M}^0_{\tilde g_1 g_3 g_4 g_5} 
= i \eta g^4 \frac{1}{\sqrt{2}} \sum_{(i,j,k) \in P_C(3,4,5)} 
\left[ (T^{a_i}T^{a_j}T^{a_k})_{i_1 i_2} - \frac{1}{N}
\delta_{i_1i_2} \mbox{Tr}(T^{a_i}T^{a_j}T^{a_k})
\right] 
M_{\tilde g g g g}^0 (p_1,p_i,p_j,p_k) \;,
\end{equation}
where the sum runs only over cyclic permutations, since the colour-ordered 
amplitudes $M_{\tilde g_1 g g g}^0$ each contain the difference of two 
colour-orderings which are inverse to each other, as shown in 
Figure~\ref{fig:d40}. It can be shown that the 
$1/N$-term in the above expression does not contribute to the 
physical scattering amplitude~\cite{ddm}.

The resulting squared matrix element, averaged over identical 
final state gluon permutations is 
\begin{eqnarray}
\frac{1}{3!}\,\left|{\cal M}^0_{\tilde g_1 g_3 g_4 g_5}\right|^2 
&=& \eta^2 g^4\, \frac{N^2-1}{16}\, \frac{1}{3!}\,  N^2
\sum_{(i,j,k) \in P_C(3,4,5)} 
\left|M_{\tilde g g g g}^0 (p_1,p_i,p_j,p_k)\right|^2 
\;.
\end{eqnarray}
It should be 
noted that this squared matrix element contains only  the leading colour
term obtained from the squares of the individual colour-ordered amplitudes,
as expected in the colour ordered formulation for a process with three 
gluons~\cite{colord,ddm}. 

The tree level amplitude for $\tilde{\chi}(q) \to \tilde{g}(p_1)
q(p_3) \bar q(p_4) g(p_5)$, contracted with $\sqrt{2} T^{a_1}_{i_1i_2} $
contains three colour structures,
\begin{eqnarray}
{\cal M}^0_{\tilde{g}_1 q_3 \bar q_4 g_5} 
&=& i \eta g^2 (-i \sqrt{2}) \Bigg[ T^{a_5}_{i_1i_4}\delta_{i_3i_2} 
M_{\tilde{g} q \bar q g}^0 (p_1,p_3,p_4,p_5) + 
T^{a_5}_{i_3i_2}\delta_{i_1i_4} 
M_{\tilde{g} q \bar q g}^0 (p_1,p_4,p_3,p_5) \nonumber \\
&& \hspace{2cm} - \frac{1}{N} T^{a_5}_{i_3i_4}\delta_{i_1i_2} 
\tilde{M}_{\tilde{g} q \bar q g}^0 (p_1,p_4,p_3,p_5) \Bigg]\;.
\end{eqnarray}
The relation between leading and subleading colour 
ordered amplitudes is
\begin{equation}
M_{\tilde{g} q \bar q g}^0 (p_1,p_3,p_4,p_5) 
+ M_{\tilde{g} q \bar q g}^0 (p_1,p_4,p_3,p_5)
=  \tilde{M}_{\tilde{g} q \bar q g}^0 (p_1,p_3,p_4,p_5)\;.
\end{equation}
The squared matrix element reads
\begin{eqnarray}
|{\cal M}^0_{\tilde{g}_1 q_3 \bar q_4 g_5} |^2
 &=& \eta^2 g^4 \left( N^2-1\right) \, N_F
\Bigg\{ N \left[ \left| 
M_{\tilde{g} q \bar q g}^0 (p_1,p_3,p_4,p_5)\right|^2
+  \left| M_{\tilde{g} q \bar q g}^0 
(p_1,p_4,p_3,p_5)\right|^2 \right] 
\nonumber \\ && \hspace{1cm}
- \frac{1}{N} \left| 
\tilde{M}_{\tilde{g} q \bar q g}^0 
(p_1,p_3,p_4,p_5)\right|^2 \Bigg\} \;.
\end{eqnarray}
\begin{figure}[t]
\begin{center}
\parbox{10cm}{\begin{picture}(10,1.5)
\thicklines
\put(0.2,0){\line(1,0){4.0}}
\put(0.2,0){\line(0,1){1.0}}
\put(4.2,0){\line(0,1){1.0}}
\put(1.15,0){\line(0,1){1.0}}
\put(1.25,0){\line(0,1){1.0}}
\put(2.15,0){\line(0,1){1.0}}
\put(2.25,0){\line(0,1){1.0}}
\put(3.15,0){\line(0,1){1.0}}
\put(3.25,0){\line(0,1){1.0}}
\put(1.2,0){\circle*{0.2}}
\put(2.2,0){\circle*{0.2}}
\put(3.2,0){\circle*{0.2}}
\put(0.1,1.2){\makebox(0,0)[l]{$i_1$}}
\put(4.1,1.2){\makebox(0,0)[l]{$i_2$}}
\put(1.1,1.2){\makebox(0,0)[l]{$a_i$}}
\put(2.1,1.2){\makebox(0,0)[l]{$a_j$}}
\put(3.1,1.2){\makebox(0,0)[l]{$a_k$}}
\put(5.2,0){\line(1,0){4.0}}
\put(5.2,0){\line(0,1){1.0}}
\put(9.2,0){\line(0,1){1.0}}
\put(6.15,0){\line(0,1){1.0}}
\put(6.25,0){\line(0,1){1.0}}
\put(7.15,0){\line(0,1){1.0}}
\put(7.25,0){\line(0,1){1.0}}
\put(8.15,0){\line(0,1){1.0}}
\put(8.25,0){\line(0,1){1.0}}
\put(6.2,0){\circle*{0.2}}
\put(7.2,0){\circle*{0.2}}
\put(8.2,0){\circle*{0.2}}
\put(5.1,1.2){\makebox(0,0)[l]{$i_2$}}
\put(9.1,1.2){\makebox(0,0)[l]{$i_1$}}
\put(6.1,1.2){\makebox(0,0)[l]{$a_i$}}
\put(7.1,1.2){\makebox(0,0)[l]{$a_j$}}
\put(8.1,1.2){\makebox(0,0)[l]{$a_k$}}
\put(4.6,0.6){\makebox(0,0)[l]{$-$}}
\end{picture}}
\end{center}
\caption{Colour flow contained in the colour ordered amplitude 
$M_{\tilde g g g g}^0 (p_1,p_i,p_j,p_k)$ contributing to the 
tree level 
decay $\tilde \chi \to \tilde g g gg$.}
\label{fig:d40}
\begin{center}
\parbox{9cm}{\begin{picture}(8.5,1.5)
\thicklines
\put(0.2,0){\line(1,0){2.0}}
\put(2.4,0){\line(1,0){0.8}}
\put(0.2,0){\line(0,1){1.0}}
\put(3.2,0){\line(0,1){1.0}}
\put(1.15,0){\line(0,1){1.0}}
\put(1.25,0){\line(0,1){1.0}}
\put(2.2,0){\line(0,1){1.0}}
\put(2.4,0){\line(0,1){1.0}}
\put(1.2,0){\circle*{0.2}}
\put(0.1,1.2){\makebox(0,0)[l]{$i_1$}}
\put(3.1,1.2){\makebox(0,0)[l]{$i_2$}}
\put(1.1,1.2){\makebox(0,0)[l]{$a_5$}}
\put(2.0,1.2){\makebox(0,0)[l]{$i_4$}}
\put(2.35,1.2){\makebox(0,0)[l]{$i_3$}}
\put(5.2,0){\line(1,0){2.0}}
\put(7.4,0){\line(1,0){0.8}}
\put(5.2,0){\line(0,1){1.0}}
\put(8.2,0){\line(0,1){1.0}}
\put(6.15,0){\line(0,1){1.0}}
\put(6.25,0){\line(0,1){1.0}}
\put(7.2,0){\line(0,1){1.0}}
\put(7.4,0){\line(0,1){1.0}}
\put(6.2,0){\circle*{0.2}}
\put(5.1,1.2){\makebox(0,0)[l]{$i_3$}}
\put(8.1,1.2){\makebox(0,0)[l]{$i_2$}}
\put(6.1,1.2){\makebox(0,0)[l]{$a_5$}}
\put(7.0,1.2){\makebox(0,0)[l]{$i_4$}}
\put(7.35,1.2){\makebox(0,0)[l]{$i_1$}}
\end{picture}}
\end{center}
\caption{Colour flow contained in the colour ordered amplitudes 
$M_{\tilde g q \bar q g}^0 (p_1,p_3,p_4,p_5)$ (left) and
$\tilde{M}_{\tilde g q \bar q g}^0 (p_1,p_3,p_4,p_5)$ (right)  
contributing to the 
tree level 
decay $\tilde \chi \to \tilde g q\bar q g$.}
\label{fig:e40}
\end{figure}

It can be seen that this neutralino decay matrix element contains the same 
colour-ordered antenna structures, displayed in 
Figure~\ref{fig:e40}, as the five-parton matrix element 
$\gamma^*\to q\bar q q' \bar q' g$~\cite{campbell},
 relevant to $e^+e^- \to 3j$ at NNLO: 
gluon ($p_5$) emission between the colour-connected pairs 
($p_1$,$p_3$) and ($p_1$,$p_4$) at leading colour, and gluon emission inside 
the ($p_3$,$p_4$) pair at subleading colour. In the latter case, the 
($p_3$,$p_5$,$p_4$) system forms a colour singlet, such that the gluon 
$p_5$ acts as a photon and $p_1$ becomes a photino which decouples 
completely from any singular limit. 

The four-parton tree-level neutralino matrix elements can be integrated 
over the tripole phase space~\cite{ggh}, thus making their infrared 
singularity structure explicit,
\begin{eqnarray}
{\cal T}^2_{\tilde g ggg}(q^2) 
&\equiv&
\int \d \Phi_{T,\tilde g ggg}\;\frac{1}{3!}
\left|{\cal M}^0_{\tilde g_1 g_3 g_4 g_5}\right|^2
\nonumber \\
&=& \left(\frac{\alpha_s}{2\pi}\right)^2\; {\cal T}^0_{\tilde g g}(q^2)
\left(q^2 \right)^{-2\e} 
  N^2 \Bigg[ 
\frac{5}{2\e^4}  + \frac{37}{4\e^3} + \frac{1}{\e^2} 
\left(\frac{398}{9} - \frac{11\pi^2}{3} \right) \nonumber \\
&&+ \frac{1}{\e} \left(
            \frac{28319}{144}
          - \frac{55\pi^2}{4}
          - \frac{188}{3}\zeta_3 \right) 
+ \left(    \frac{2201527}{2592}
          - \frac{529\pi^2}{8}
          - \frac{722}{3}\zeta_3
          + \frac{511\pi^4}{720} 
\right)
+ {\cal O}(\e) \Bigg] \; , \\
{\cal T}^2_{\tilde g q\bar q g}(q^2) 
&\equiv&
\int \d \Phi_{T,\tilde g q\bar q g}\;
\left|{\cal M}^0_{\tilde g_1 q_3 \bar q_4 g_5}\right|^2
\nonumber \\
&=&  \left(\frac{\alpha_s}{2\pi}\right)^2\;{\cal T}^0_{\tilde g g}(q^2)
\left(q^2 \right)^{-2\e} 
  \Bigg[ N N_F \Bigg( 
-\frac{5}{6\e^3}  - \frac{17}{4\e^2} + \frac{1}{\e} 
\left(-\frac{2239}{108} + \frac{5\pi^2}{4} \right)
\nonumber \\ && \hspace{3.8cm}
+  \left(
  -\frac{20521}{216}
          + \frac{51\pi^2}{8}
          + \frac{200}{9}\zeta_3 \right) 
\Bigg) \nonumber \\
&& + \frac{N_F}{N} \Bigg( \frac{1}{6\e^3} + \frac{35}{36\e^2} 
+ \frac{1}{\e} \left( \frac{1045}{216}
          - \frac{\pi^2}{4}
\right) + \left(\frac{28637}{1296}
          - \frac{35\pi^2}{24}
          - \frac{40}{9}\zeta_3
\right)
\Bigg)
+ {\cal O}(\e) \Bigg] \; .
\end{eqnarray}

The sum of all NNLO subtraction terms yields the following infrared
pole structure, which can be expressed in terms of NNLO infrared 
singularity operators~\cite{catani},
\begin{eqnarray}
\lefteqn{\Poles\left({\cal T}^2_{\tilde g gg}(q^2) + 
{\cal T}^2_{\tilde g q\bar q}(q^2) + {\cal T}^2_{\tilde g ggg}(q^2) 
+ {\cal T}^2_{\tilde g q\bar q g}(q^2)\right) } \nonumber \\
&=& 
\left(\frac{\alpha_s}{2\pi}\right)^2\;{\cal T}^0_{\tilde g g}(q^2)
\left(q^2 \right)^{-2\e} 
  \Bigg[ N^2 \Bigg( -\frac{2}{\e^4} - \frac{113}{12\e^3} + \frac{1}{\e^2}
\left( -\frac{27}{4} + \frac{9\pi^2}{4} \right)
+ \frac{1}{\e} \left(
  \frac{1049}{432}
          + \frac{11\pi^2}{3}
          + \frac{28}{3}\zeta_3 \right)  \Bigg) \nonumber \\
&& + N N_F \Bigg( 
\frac{7}{6\e^3}  + \frac{61}{36\e^2} + \frac{1}{\e} 
\left(-\frac{55}{54} - \frac{5\pi^2}{12} \right)
\Bigg) + \frac{N_F}{N} \Bigg( \frac{1}{8\e} 
\Bigg) 
+ N_F^2 \Bigg( - \frac{1}{9\e^2} \Bigg) 
+ {\cal O}(\e^0) \Bigg] \; \\
&=& - \left(\frac{\alpha_s}{2\pi}\right)^2\; \mbox{Re}\Bigg[ 
  - 2{\bom I}_{qg}^{(1)}(\e,q^2) 
\left(2{\bom I}_{qg}^{(1)}(\e,q^2) +2{\bom I}_{qg}^{(1),*}(\e,q^2)
 \right){\cal T}^0_{\tilde g g}(q^2)
  -2 \frac{\beta_0}{\e}  
\, 2{\bom I}_{qg}^{(1)}(\e,q^2) {\cal T}^0_{\tilde g g}(q^2)
 \nonumber\\
&&\hspace{2.4cm}
+ 4 \,  {\bom I}_{qg}^{(1)}(\e,q^2) {\cal T}^1_{\tilde g g}(q^2)
+ 2 \,
e^{-\e\gamma } \frac{ \Gamma(1-2\e)}{\Gamma(1-\e)} 
\left(\frac{\beta_0}{\e} + K\right)\, 2
 {\bom I}_{qg}^{(1)}(2\e,q^2) {\cal T}^0_{\tilde g g}(q^2) \nonumber\\
&& \hspace{2.4cm}
+ 2 \, {\bom H}_{\tilde g g}^{(2)}(\e,q^2){\cal T}^0_{\tilde g g}(q^2)
\, \Bigg] ,
\label{eq:I2}
\end{eqnarray}
where $\beta_0$ is the first term of the QCD $\beta$-function
(\ref{eq:qcdbeta}) and the constant $K$ 
\begin{equation}
K = \left( \frac{67}{18} - \frac{\pi^2}{6} \right) N - 
\frac{5}{9}  N_F.
\end{equation}
This structure coincides precisely with the singularity structure 
predicted in~\cite{catani} for the purely virtual (two-loop times tree plus 
one-loop self-interference) NNLO corrections to a tree level process 
containing {\it two} quark-gluon antenna functions. The final state dependent 
constant ${\bom H}_{\tilde g g}^{(2)}(\e,q^2)$ contributes only at 
${\cal O}(\e^{-1})$:
\begin{equation}
{\bom H}_{\tilde g g}^{(2)}(\epsilon,q^2)
=\frac{e^{\epsilon \gamma}}{4\,\epsilon\,\Gamma(1-\epsilon)} 
\,\left( H_{\tilde g}^{(2)} + H_g^{(2)}  \right)  \left(-q^2 \right)^{-2\e} \;.
\end{equation}
It can be related to the known constants determining the 
$\e^{-1}$ poles of 
four-parton
two-loop matrix elements involving $i$ quarks and $j$ gluons~\cite{twol}:
\begin{equation}
\label{eq:htwo}
{\bom H}_{iq,jg}^{(2)}(\epsilon,q^2)
=\frac{e^{\epsilon \gamma}}{4\,\epsilon\,\Gamma(1-\epsilon)} 
\, \left(i H_q^{(2)} + j H_g^{(2)}  \right)  \left(-q^2 \right)^{-2\e} 
\end{equation}
with 
\begin{eqnarray}
H^{(2)}_g &=&  
\left(\frac{1}{2}\zeta_3+{\frac {5}{12}}+ {\frac {11\pi^2}{144}}
\right)N^2
+{\frac {5}{27}}\,\NF^2
+\left (-{\frac {{\pi }^{2}}{72}}-{\frac {89}{108}}\right ) N \NF 
-\frac{\NF}{4N}, \nonumber\\
H^{(2)}_q &=& \frac{N^2-1}{N^2}\Bigg[
\left({1\over 4}\zeta_3+{41\over 108}+{\pi^2\over 96}\right) N^2
+\left({3\over 2}\zeta_3+{3\over 32}-{\pi^2\over 8}\right)\left(N^2+1\right)
+\left({\pi^2\over 48}-{25\over 216}\right)N N_F\Bigg]\;, \nonumber \\
H^{(2)}_{\tilde g} &=&
\left({1\over 4}\zeta_3+{41\over 108}+{\pi^2\over 96}\right) (2N^2)
+\left({\pi^2\over 48}-{25\over 216}\right){2(N^2-1)N_F\over N}
-  \left(\frac{13}{14} - \frac{\pi^2}{8} + \frac{1}{2} 
\zeta_3\right)\, (2 N^2)\;.
\label{eq:H2}
\end{eqnarray}
In these equations, we decomposed the $H^{(2)}_i$ according to their colour 
structures. The coefficient $(N^2+1)$ in front of the subleading colour 
contribution to $H^{(2)}_q$ arises due to the fact that abelian 
diagrams contributing to $q ggg\bar q$ final states carry this colour 
structure, such that the generic (planar) leading colour contribution is 
given by just the first term in  $H^{(2)}_q$ (see also Equation (3.6) 
of~\cite{campbell}). 

It can be seen that $H^{(2)}_{\tilde g}$ contains twice the leading colour 
and the flavour dependent terms of  $H^{(2)}_{q}$. 
The subleading colour term is absent, 
and the last term can be identified with the contribution to 
$H^{(2)}_{q}$ from singularities arising from final states containing 
two quark-antiquark pairs of identical flavour 
(Equation (4.51) of~\cite{our2j})\footnote{In neutralino decay, this 
identical-flavour contribution 
corresponds to final states with three gluinos and one gluon. Being a 
SUSY QCD correction, it is discarded here, and has to be subtracted from 
$H^{(2)}_{\tilde g}$.}.

Equations~(\ref{eq:I2}) and (\ref{eq:H2}) demonstrate that 
the NNLO three and four parton contributions to neutralino decay into a 
gluino and massless partons display the same singularity structure as 
final state observables containing adjacent quark-gluon pairs, 
provided that the colour factors are adjusted correctly. 
It is therefore possible 
 to construct colour-ordered quark-gluon antenna functions 
from the  neutralino decay matrix elements derived here from the effective 
Lagrangian density (\ref{eq:hw}).

\section{Conclusions and Outlook}
\setcounter{equation}{0}

QCD antenna functions describe the behaviour of QCD matrix elements in  their
infrared singular limits, corresponding to soft or collinear parton emission.
They are constructed so that they describe all singular limits arising  from
emission of unresolved partons in between the two colour-connected
hard partons that define the  antenna.  The
quark-antiquark antenna function is directly related to the physical matrix
elements for $\gamma^* \to q\bar q +$gluons.  However,  up to now, the  NLO
quark-gluon and gluon-gluon antenna functions~\cite{cullen}
were  constructed by starting
from the quark-antiquark antenna function and adding terms to match the
remaining limits contained in them. It does not appear feasible to extend this
procedure to higher orders.  

In this paper, we demonstrated that quark-gluon 
QCD antenna functions to all orders 
can be derived from an effective Lagrangian 
describing the decay of a massive neutralino into a massless gluino and 
the gluon field. In the colour ordered formalism underlying the antenna 
functions, 
the Majorana nature of the gluino allows it to appear simultaneously as
quark and as antiquark. 
We demonstrated that the physical neutralino decay matrix elements
reproduce the singular structure of QCD quark-gluon antenna functions 
at NLO and NNLO. We extracted the infrared structure for decay kinematics, 
as required for jet observables without partons in the initial state. By 
analytic continuation, the matrix elements derived here can also be continued
to production (leading order process contains 
partons only in the initial state) or scattering (leading order 
process contains partons in initial and final state) kinematics, 
where they have to be integrated over the appropriate phase spaces.

All QCD antenna functions can be derived 
(as opposed to constructed) from physical matrix elements: quark-antiquark 
antennae from the decay of a virtual photon into partons, quark-gluon antennae
from neutralino decay into gluino plus partons and finally gluon-gluon
antennae~\cite{hnew} from Higgs boson decay into partons through the effective 
Lagrangian~\cite{hgg}
coupling the Higgs field to the gluonic field strength tensor. 
The NNLO antenna subtraction functions obtained through this procedure will be 
reported in  a subsequent publication~\cite{gggsub}.

\section*{Acknowledgements} 
The authors would like to thank Werner Porod and Daniel Wyler for numerous
discussions on neutralino interactions. 

This research was supported in part by the Swiss National Science Foundation 
(SNF) under contracts PMPD2-106101  and 200021-101874, 
 by the UK Particle Physics and Astronomy  Research Council and by
the EU Fifth Framework Programme  `Improving Human Potential', Research
Training Network `Particle Physics Phenomenology  at High Energy Colliders',
contract HPRN-CT-2000-00149.

\end{document}